\documentstyle[prl,aps,preprint,tighten,floats,aps,epsf,psfig]{revtex}
\def\simlt{\stackrel{<}{{}_\sim}}

\def\be{\begin{equation}}
\def\ee{\end{equation}}
\def\bear{\begin{eqnarray}}
\def\eear{\end{eqnarray}}
\def\beqn{\begin{eqnarray}}
\def\eeqn{\end{eqnarray}}

\def\simlt{\stackrel{<}{{}_\sim} }

\begin{document}
\draft
\preprint{\vbox{\baselineskip=12pt
\rightline{FERMILAB-PUB-99/230-T}
\vskip0.2truecm
\rightline{hep-ph/9908326}}}

\title{Superstring Theory and CP- Violating Phases: Can They Be
Related?} 
\author{
M. Brhlik${}^{\dagger}$, L. Everett${}^{\dagger}$, G. L.
Kane${}^{\dagger}$,
and J. Lykken${}^*$}
\address{${}^{\dagger}$ Randall Laboratory, Department of Physics,  
University of Michigan\\
         Ann Arbor, Michigan, 48109, USA \\
${}^*$Theoretical Physics Department, Fermi National Accelerator
Laboratory\\
Batavia, Illinois, 60510, USA}
\maketitle
\begin{abstract}
We investigate the possibility of large CP- violating phases in
the soft breaking terms derived in superstring models. The
bounds on the electric dipole moments (EDM's) of the electron and neutron
are satisfied through cancellations occuring because of the structure of
the string models.
Three general classes of four-dimensional string models are considered:
(i) orbifold compactifications of perturbative heterotic string theory,
(ii) scenarios based on Ho\v{r}ava-Witten theory,
and (iii) Type I string models (Type IIB orientifolds).
Nonuniversal phases of the gaugino
mass parameters greatly facilitate the necessary cancellations among the
various contributions to the EDM's;
in the overall modulus limit, the gaugino masses are universal at tree
level in both the perturbative
heterotic models and the Ho\v{r}ava-Witten scenarios,
which severely restricts the
allowed regions of parameter space. Nonuniversal gaugino masses do arise
at one-loop in the heterotic orbifold models, providing for corners of
parameter space with ${\cal O}(1)$ phases consistent with the
phenomenological bounds.
However, there is a possibility of nonuniversal gaugino masses at
tree level in the Type I models, depending on the details of the
embedding of the SM into the D- brane sectors.  We find that
in a minimal model with a particular embedding of the Standard Model gauge
group into two D- brane sectors, viable large phase solutions can be
obtained over a wide range of parameter space.

\end{abstract}
\newpage
\section{Introduction}

A central issue to be addressed in supersymmetric theories  
is the origin and dynamical mechanism of spontaneous supersymmetry
breaking. In supersymmetric extensions of the Standard Model (SM) such as
the Minimal Supersymmetric Standard Model (MSSM),  the effects of
the unknown dynamics of supersymmetry breaking are encoded by adding
terms to the Lagrangian which break supersymmetry explicitly; these
terms depend on a considerable number of parameters which can be
considered as independent in the phenomenological analysis of the model.
For example, the most general set of soft supersymmetry breaking
parameters in the MSSM, which is defined to be the
minimal supersymmetric extension of the SM with the standard Higgs sector
and conserved R-parity, includes 105 masses,
mixing angles, and phases (not counting the gravitino mass
and coupling) \cite{dimopoulos}. From the phenomenological point
of view, this large number of parameters can be cumbersome but not
otherwise problematic, as it is for experiments to measure and for the
underlying theory (for example, superstring theory) to predict the values
of these parameters. Phenomenological analyses aid in this process both
for experimentalists and  theorists by serving as a helpful guide to the
allowed regions of parameter space, and are crucial from the experimental
side since almost none of the Lagrangian parameters are directly measured.

Due to the large number of parameters of the soft
breaking Lagrangian, restricted sets of parameters are often chosen to
simplify the analysis. While this approach is sensible, it is important
not to exclude possibly allowed regions of parameter space based on
potentially misleading theoretical assumptions.   An example is the
conventional
statement of the supersymmetric CP problem, which is that the CP-
violating phases in the MSSM (which arise in the soft breaking Lagrangian
and in the phase of $\mu$) are individually constrained to be less than
${\cal O}(10^{-2})$ for sparticle masses at the TeV scale by the
experimental upper limits for the electric dipole moments of the electron
and the neutron \cite{oldedm,edm,gar}.  Based on this argument, these CP-
violating phases have traditionally been set to zero in phenomenological
analyses. 

However, a recent reinvestigation of this issue \cite{nath,bgk}, see also
\cite{pokorski} has demonstrated that cancellations between different
contributions to the electric dipole moments can allow for regions of
parameter space with phases of ${\cal O}(1)$ and light sparticle masses
that satisfy the
phenomenological constraints, contrary to conventional wisdom.  If these
phases are in fact nonzero (which future experiments will need to
determine), they can have important effects on many physical observables,
and thus on the extraction of the values of the soft supersymmetry
breaking parameters from experimental measurements \cite{bk}. A thorough
numerical analysis including the seven significant phases \cite{bgk}
indicates that the cancellations can only occur for the large phase
solutions if the various soft breaking parameters satisfy particular
approximate relations. Such relations may provide clues to the dynamical
mechanism of supersymmetry breaking, and hence to the form of the
underlying theory.  

As superstring theory is the best candidate for the underlying fundamental
theory of all interactions, it is desirable 
to investigate the phase structure of soft supersymmetry breaking
terms that can arise in classes of four-dimensional superstring models.
In this paper, we address the question of whether the relations among the
soft breaking parameters derived in these models allow for large phases
that satisfy the electric dipole moment constraints via the
cancellations. 

CP is a discrete gauge symmetry in string theory, and thus can only
be broken spontaneously \cite{dine}. If this breaking occurs via the
dynamics of compactification and/or supersymmetry breaking, then the
four dimensional effective field theory will exhibit explicit
CP- violating phases.
The origin of supersymmetry breaking in string theory
remains an unresolved issue, though it is known to be
nonperturbative. However, progress in addressing the low
energy implications of supersymmetry breaking in string theory can be made
by utilizing the phenomenological approach of Brignole, Ib\'a\~nez, and
Mu\~noz\cite{bim}. In this approach, the degrees of freedom involved in
supersymmetry breaking are assumed to be the dilaton $S$ and (untwisted)
moduli $T_m$, which are superfields generically present in
four-dimensional string models. The
effects of the unknown nonperturbative dynamics that break supersymmetry
are then encoded in the $F$- component VEV's of these superfields.  These
VEV's are conveniently parameterized in terms of Goldstino angles, which
denote the relative contributions of each field to the supersymmetry
breaking process.  
The phases (if nonzero) of the  $F$- component VEV's of the dilaton and
moduli provide the main sources for CP- violating phases in the soft
terms.  Whether or not these VEV's have sizeable phases is a dynamical
question that cannot be addressed within this framework, and hence these
phases are treated as independent parameters in the analysis. It is
important to note that to obtain the traditional resolution to the
supersymmetric CP problem in a natural manner,  dynamical
principles are required which guarantee that not only the phases of the
$F$- component VEV's but also the phase of $\mu$ (which in principle
has a different origin) are zero or negligibly
small.  While arguments for such principles exist (primarily within the
context of perturbative heterotic string models), our strategy has been
that it will be experimental information that is likely to play an
important role in determining or constraining the values of these
parameters.

Within particular classes of four-dimensional string models, the couplings
of the dilaton, moduli, and matter fields are calculable, which leads in
turn to a specific pattern of soft breaking parameters at the string
scale (as a function of the unknown $F$-component VEV's, which serve as
input parameters). In our analysis of the phase structure of the soft
breaking terms, we use the renormalization group equations (RGE's) to
obtain the values of the parameters at the electroweak scale and
subsequently compute
the EDM's of the electron and neutron.  In the analysis, it is
important to note that the general results of \cite{bgk} illustrate that 
sufficient cancellations among the various contributions to
the EDM's are difficult to achieve unless there are large relative phases
in the gaugino masses.  Furthermore, the phases of the gaugino mass
parameters do not run at one-loop order, and thus at the electroweak scale
only deviate from their string-scale values by small two-loop corrections.
Therefore, the possibility of viable large phase solutions crucially
depends on whether nonuniversal gaugino mass parameters are predicted
within a given string model. 

Following \cite{bim,bims,bimrev}, we first consider models derived from
weakly coupled heterotic string theory in orbifold
compactifications\cite{orbifolds}, and focus on the case in which $S$ and
one ``overall modulus" field $T$ participate in the supersymmetry
breaking. In these models, the dilaton
$F$ term leads to universal gaugino masses at tree-level; however, the
modulus field $T$ does provide a nonuniversal contribution at one-loop
order (although loop-suppressed). We thus focus on a generalized version
of their $O-II$ scenario, with arbitrary phases for $S$ and $T$, and
consider the moduli-dominated limit.  We find that the phase structure of
the soft terms does allow for small regions in the parameter space for
which the EDM constraints can be satisfied with large phases, and that the
results may depend on the particular (model-dependent) solution employed
for the $\mu$ problem. This fact is very encouraging, as it suggests such
analyses may help us learn how the $\mu$ problem is solved.

We next consider the soft breaking terms which arise in newer classes of 
four-dimensional string models,  including the Ho\v{r}ava-Witten
scenarios, and models within the general perturbative
Type I string
picture.  In contrast to the weakly coupled heterotic models, the
calculational and model-building techniques in each of these scenarios are
at early stages, and there is as yet no quasi-realistic model.  However,
recent studies have indicated that the phenomenological properties of
these classes of models, including the patterns of the soft supersymmetry
breaking parameters, can be quite distinctive from those of the
perturbative heterotic models traditionally studied in superstring
phenomenology. 

We first consider models based on the Ho\v{r}ava-Witten theory
\cite{horavawitten} (11-dimensional supergravity compactified on a
Calabi-Yau manifold times the eleventh segment) in which the observable
sector gauge groups arise from the $E_8$ gauge group on one of the
ten-dimensional boundaries, for which the soft terms of the effective
supergravity theory \cite{ovrut} have been computed \cite{nilles,munoz}.  
In contrast to the perturbative heterotic case, both
the dilaton and modulus fields contribute to the gaugino masses with
${\cal O}(1)$ coefficients. However, in the limit in which only the
overall modulus $T$ contributes to SUSY breaking, the gaugino masses are
universal in this scenario. Hence, significant CP- violating phases in
the soft terms consistent with the phenomenological bounds are disallowed
over the majority of parameter space (the situation is analogous to that
of the dilaton-dominated limit of perturbative heterotic orbifold models
studied in \cite{barrkhalil}).

However, within the more general Type I string picture
\cite{sagnotti,kakushadze,ibanez,shiu},
there is the possibility of nonuniversal gaugino masses at tree level,
which has important implications for the possible CP- violating effects.
As an illustrative example of models within this framework, we focus on
the four-dimensional Type IIB orientifold models, in which consistency
conditions (tadpole cancellation) require the addition of open string
(Type I) sectors and Dirichlet branes, upon which the open strings must
end.  The patterns of the soft supersymmetry breaking terms arising in
this class of models crucially depend on the embedding of the SM gauge
group into the D- brane sectors.
In particular, nonuniversal gaugino masses can be obtained at
tree-level if the SM gauge groups are embedded in two distinct D- brane
sectors, in direct contrast to the perturbative heterotic
orbifolds and the Ho\v{r}ava-Witten scenarios described above.
Our analysis indicates that within a minimal model in which $SU(3)$ and
$U(1)_Y$ (but not $SU(2)$) arise from the same D- brane sector, the
necessary cancellations between the contributions to the EDM's 
 occur over a wide range of parameter
space.  The results of this study illustrate that
as viable large soft phases depend on how the SM is embedded and how the
$\mu$ problem is solved, etc., we may be able to learn about (possibly
nonperturbative) Planck scale physics using low energy data.

This paper is structured as follows.  In Section \ref{II}, 
we briefly review the method and results of the general EDM 
calculation of \cite{nath,bgk}, with an
emphasis on issues relevant for our analysis. We present the analysis of
the soft breaking terms from the perturbative heterotic orbifold models in
Section \ref{III}.  In Section \ref{IV}, we consider the
Ho\v{r}ava-Witten scenarios, and in Section \ref{V}, we
analyze the Type I models.  Finally, we present the summary
and conclusions in Section \ref{VI}.
  
\section{Electric Dipole Moment Calculation}
\label{II}
The purpose of this paper is to examine whether the (complex)
soft breaking parameters which can arise in classes of string-derived
models can satisfy the phenomenological bounds from the EDM's by
cancellations.  We build
upon the recent calculations of the electric dipole moments presented in
\cite{nath} and \cite{bgk}. In this section, we briefly summarize 
the framework and results of these calculations, and comment upon the
issues to be addressed in our analysis of string-motivated models of
the soft breaking parameters.

It is well known that in (softly broken) supersymmetric theories with
CP- violating phases of ${\cal O}(1)$, superpartner exchange at the
one-loop level can lead to contributions to the electric dipole moments of
the fermions which can exceed the experimental upper bounds 
\cite{oldedm,edm,gar}. As previously mentioned, the traditional resolution
to this problem has been to constrain the phases to be less than ${\cal
O}(10^{-2})$ (which can be 
interpreted as fine-tuning), or assume heavy sfermion masses (which can
violate naturalness).  However, the issue was reinvestigated
first by \cite{nath}, and subsequently in \cite{bgk}, in which
the EDM's were computed using an effective theory approach in which the
contributions from chargino, neutralino, and gluino loops to the
relevant Wilson coefficients were determined numerically. In their
work, the main emphasis was on the possibility of cancellations between
the various contributions to the Wilson coefficients. This mechanism can
allow large values of the phases to give
contributions consistent with the
experimental bounds on the values of the electric dipole moments $d_e$,
$d_n$ of the electron and the neutron, respectively.  The current
limits for the neutron EDM require that \cite{nexpnew}
\begin{eqnarray}
|d_n|< 6.3 \times 10^{-26} {\ \rm ecm},
\end{eqnarray}
at 90\% confidence level, and for the electron EDM \cite{eexp}
\begin{eqnarray}
|d_e|< 4.3 \times 10^{-27} {\ \rm ecm},
\end{eqnarray}
at 95\% confidence level.

In \cite{bgk}, a general set of CP- violating phases is assumed. For
simplicity, the phases of the off-diagonal terms in the scalar mass
matrices are neglected, as the impact of these phases on physical
observables may be suppressed by the same mechanism required to suppress
FCNC\footnote{For our purposes, this statement indicates that we do not
consider K\"{a}hler potentials with off-diagonal metric; for further
discussions of this issue see e.g. \cite{bim,bims,bimrev}.}, and in any
case are unlikely to modify the results qualitatively.  
The phases included in the analysis are thus the phases of the
gaugino masses $M_{1,2,3}$, the phases of the $\mu$ term and the
associated $b=B\mu$ parameter, and the phases of the
$A$ parameters associated with the trilinear scalar couplings.
However, as noted in \cite{thomas,bgk,wagner,reparameterization} and
references therein, not all of these phases are physical due to additional
approximate global $U(1)$ symmetries of the MSSM Lagrangian which can be
promoted to full symmetries by treating the parameters as spurions charged
under those symmetries. The result is that there is the freedom to
rotate away one of the phases in the gaugino mass sector and also to set
$b$ (and the VEV's of the Higgs doublets) to be real at the
electroweak scale without loss of generality.  The phase $\varphi_2$ of
the $SU(2)$ gaugino mass $M_2$ is set to zero in the parameterization
choice of \cite{bgk}, and hence the relevant  phases in the analysis are
$\varphi_1$, $\varphi_3$ of the $U(1)_Y$ and $SU(3)$ gaugino masses,
$\varphi_{\mu}$, $\varphi_{A_u}$, $\varphi_{A_d}$, $\varphi_{A_t}$ and
$\varphi_{A_e}$ (in self-evident notation). Note that in minimal
supergravity-inspired models as studied in \cite{nath}, the gaugino masses
can be taken to be real without loss of generality, and then there are
only two relevant phases, a common $\varphi_{A}$ and $\varphi_{\mu}$.

As $\mu$ and $B$ (and their phases) are
relevant in the analysis,  the results will in general depend
on the solution to the $\mu$ problem. In string models, the ``bare" $\mu$
term is absent in the superpotential,
since the fields are massless at the string scale, and there are several
possibilities for the generation of an effective $\mu$ term (either in the
superpotential or in the K\"{a}hler potential \cite{muproblem}) without
invoking additional
gauge singlet matter fields; we refer the reader to \cite{bim,orbimu} for 
further discussions of this issue. The results
for the phase of $\mu$ and that of the associated $B$ term strongly
depends on which solution (or in fact if both mechanisms are present) is
preferred in a given model.  Since these issues are highly
model-dependent, an additional possibility is to treat the phases of $\mu$
and $B$ as independent parameters; their magnitudes are naturally
constrained by the requirement of correct electroweak symmetry breaking.
This sensitivity to the way $\mu$ is generated is a very positive feature,
since it implies that data on the phases may help determine experimentally
how $\mu$ is generated.

The general results of \cite{bgk} demonstrate that
sufficient cancellations among the various contributions to
the EDM's are difficult to achieve unless there are large relative phases
in the soft masses of the gaugino sector. This feature is due to the
approximate $U(1)_R$ symmetry of the Lagrangian of the MSSM \cite{thomas},
which allows one of the phases of the gaugino masses to be set to zero at
the electroweak scale without loss of generality
\cite{thomas,bgk,wagner,reparameterization}.
Furthermore, the phases of the gaugino mass parameters do not run at
one-loop order, and thus at the electroweak scale only deviate from the
string-scale values by small two-loop corrections. Therefore, if the
phases of the gaugino masses are universal at the string scale, they will
be approximately zero at the electroweak scale (after the $U(1)_R$
rotation).  Cancellations among the chargino and neutralino
contributions to the electron EDM are then necessarily due to the
interplay between the phases of $A_e$ and $\mu$.
The analysis of \cite{bgk} demonstrates that cancellations are then
difficult to achieve  as the pure gaugino part of the neutralino diagram
adds destructively with the contribution from the gaugino-higgsino mixing,
which in turn has to cancel against the chargino
diagram. As a result, the cancellations are generally
insufficient, and hence in this case over most of the parameter
space the phases of the other soft breaking parameters as well as the
$\mu$ parameter must naturally be $\simlt  10^{-2}$ (the traditional
bound) \cite{edm} unless the sfermion masses are greater than ${\cal
O}(\rm{TeV})$~\footnote{We note that this situation is precisely that of
the minimal supergravity case studied in \cite{nath}, as the gaugino
masses are universal in this scenario. In this case, a heavy superpartner
spectrum is required to exhibit regions of parameter space with
large phases consistent with
the phenomenological bounds on the EDM's.}.

Therefore, the possibility of large CP- violating phases in the 
string-motivated models of soft breaking terms we consider depends
significantly on whether the gaugino masses are allowed to have large
relative phases ({\it i.e.}, if they are nonuniversal).  It is important 
to note that the gauginos can be degenerate (or nearly degenerate) in mass
at the string scale and have different phases; in practice, we find
examples where this holds.  In the next sections, this feature will be
displayed explicitly in the analysis of the soft breaking parameters in
three classes of four-dimensional string models.

\section{Soft Breaking Terms in Perturbative Heterotic Superstring Models}
\label{III}

\subsection{Theoretical Framework}

In the analysis of \cite{bim,bims}, the primary assumption is that
supersymmetry is broken by a combination of the dilaton field $S$ and the
moduli $T_m$ present in generic four-dimensional string models.   These
fields have a vanishing (perturbative) scalar potential and
gravitationally suppressed interactions with the fields of the observable
sector, and thus are natural candidates to play a role in the breakdown of
supersymmetry.   

In classes of four-dimensional models derived from
perturbative heterotic superstring theory, the  K\"{a}hler potential
$K$, gauge kinetic function
$f_a$ (where $a$ labels the gauge groups), and superpotential $W$ are
calculable (generally to one-loop order~\footnote{Due to the
holomorphicity of the
superpotential and the (Wilsonian) gauge kinetic function,
nonrenormalization theorems imply these functions do not receive
higher-loop corrections.  However, the K\"{a}hler potential does
receive loop corrections, and thus is the least well-determined function
of the string theory effective action.  For further details, see
\cite{quevedo} and references therein.}) in string perturbation
theory.  The calculational techniques have been particularly well
developed for the case of orbifold compactifications
(see for example \cite{orbifolds,quevedo}).  However, the nonperturbative
contributions to the K\"{a}hler potential and the superpotential, which
play a crucial role in supersymmetry breaking, remain uncertain.
In the absence of the knowledge of how supersymmetry is broken, the
authors of \cite{bim} proposed an efficient parameterization of the soft
breaking terms in terms of the (unknown) $F$- component VEV's of $S$ and
$T_m$.  For example, for the case in which the fields which break
supersymmetry are just $S$ and the ``overall modulus" $T$ associated with
the radius of the compactification manifold~\footnote{As in \cite{bim}, we
consider the overall modulus case both for simplicity and because the $T$
modulus is always present in generic four-dimensional
string models. We comment later about the implications for the purposes of
this study of relaxing this assumption, and refer the reader to
\cite{bims} for a discussion of the multimoduli case in the
perturbative heterotic orbifold models.},
the $F$- component VEV's can be expressed as follows (assuming no mixing
among their kinetic terms):
\begin{eqnarray}
\label{fvevs0}
F^S&=&\sqrt{3}m_{3/2}(S+S^{*})e^{i\alpha_S}\sin\theta\nonumber\\
F^T&=&m_{3/2}(T+T^{*})e^{i\alpha_T}\cos\theta,
\end{eqnarray}
in which $m_{3/2}$ is the gravitino mass and 
$\alpha_S$, $\alpha_T$ denote the (in this parameterization arbitrary)
phases.  $\theta$ is the Goldstino angle, which measures the relative
contributions of $S$ and $T$ to the supersymmetry breaking; the $\sin
\theta \rightarrow 1$ and  $\sin \theta \rightarrow 0$ limits correspond
to dilaton and moduli dominance, respectively.  
The soft terms in the case of general orbifold models have been computed
in \cite{bim}.  The results demonstrate the advantage of
the parameterization (\ref{fvevs0}), as the soft terms take on very simple
forms when expressed in terms of these parameters.

We note in passing that in specific scenarios for spontaneous
supersymmetry breaking such as gaugino condensation in the hidden sector,
the form of the nonperturbative superpotential $W_{np}(S,T)$ is known and
the values of $F^{S}$, $F^{T}$ can in principle be determined.  However,
explicit models typically suffer from generic problems, including that
of the runaway dilaton and a nonvanishing, negative cosmological constant. 
We choose to follow \cite{bim} and consider the parameters in
(\ref{fvevs0}) as free parameters, allowing in particular for nonzero
values of $\alpha_S$,
$\alpha_T$. Our philosophy is that experimental information will determine
or constrain the parameters, thereby leading theorists to
recognize how supersymmetry is broken.  We comment briefly below on the
types of parameter ranges for $\alpha_S$, $\alpha_T$ encountered in the
gaugino condensation scenarios,
and refer the reader to \cite{bim,bimrev} and references therein for more
comprehensive discussions.

As discussed in the previous section, large CP effects
consistent with the phenomenological bounds on the EDM's
generally require large relative phases in the gaugino mass
parameters, which implies nonuniversal gaugino masses. In general, the
source for the gaugino masses is the (field-dependent) gauge kinetic
function $f(S,T)$, which in perturbative heterotic string theory is
independent of $T$ at tree level 
and given by 
\begin{equation}
\label{ftree}
f_{a\,tree}=k_a S,
\end{equation}
in which $k_a$ is the Ka\v{c}-Moody level of the gauge group.  This
expression yields universal gaugino masses, since the dilaton couples
universally to all gauge groups.  However, the one-loop (threshold)
corrections to the gauge kinetic function have been computed in orbifold
models, and provide for the possibility of nonuniversal gaugino masses.
These corrections depend on $T$ as follows:
\begin{equation}
\label{f1loop}
f_{a\,1-loop}=-\frac{1}{16\pi^2}(b'_a-k_a\delta_{GS})\log [\eta(T)]^4,
\end{equation}
in which $b'_a$ is a numerical coefficient dependent upon the matter
content of the model, $\eta (T)$ is the Dedekind function, 
and $\delta_{GS}$ is a coefficient (a negative integer in most orbifold
models) related to the cancellation of duality anomalies in the theory.  
This coefficient also is important in that it measures the amount of
mixing between the kinetic terms of the $S$ and $T$ fields, which
occurs in the loop corrections to the K\"{a}hler potential.

In the dilaton-dominated limit ($\sin \theta \rightarrow 1$), the
contributions to the gaugino masses from (\ref{ftree}) dominate the
one-loop corrections from (\ref{f1loop}).  Thus, the gaugino masses are  
universal in this limit and hence the phases in the gaugino sector can be
taken to vanish without loss of generality.  It is therefore unlikely that
sufficient cancellations will occur in this limit except at exceptional
points in the parameter space depending on the (model-dependent)
solution to the $\mu$ problem, and thus the traditional solutions to the
supersymmetric CP problem of either small ${\cal O} (10^{-2})$ phases or
heavy squark masses must be invoked to avoid electric dipole moments for
the electron and the neutron which violate the experimental bounds.  
The analysis of the EDM constraints within the dilaton-dominated
scenario  has recently been presented in \cite{barrkhalil}; their results 
demonstrate explicitly that large phases are disallowed over the majority
of the parameter space.

Therefore, we are naturally led to consider
the moduli-dominated ($\sin \theta \rightarrow 0$) limit and to include
one-loop corrections to $f$ (and $K$, for consistency) to obtain the
possibility of nontrivial phases for the gaugino masses.  The $S-T$ mixing
in the K\"{a}hler potential requires a slight modification of the
parameterization (\ref{fvevs0}) of the $F$- component
VEV's, which amounts to a redefinition of $\alpha_{S}$, $\alpha_{T}$, and
$\theta$; a thorough discussion
of this issue is given in \cite{bim}, to which we refer the reader for
details.

An example of this type was presented in \cite{bim} as the $O-II$
scenario, which is a moduli-dominated scenario in which the one-loop
mixing between $S$ and $T$ is crucial to avoid the vanishing of the soft
mass-squares of the scalar fields in the $\sin \theta \rightarrow 0$
limit. 
In this model, the $A$ terms and scalar mass-squares are universal, while 
the gaugino masses are nonuniversal.  The soft breaking parameters take
the form: 
\begin{eqnarray}
\label{soft1}
m_i^2&=&m^2_{3/2}(-\delta_{GS})\epsilon^{'}\nonumber\\
A_{t,e,u,d}&=&-\sqrt{3}m_{3/2}e^{-i\alpha_S} \sin \theta,
\end{eqnarray}
\begin{eqnarray}
\label{gaugmass1}
M_3&=&\sqrt{3}m_{3/2}[e^{-i\alpha_S}\sin \theta -
(3+\delta_{GS})\epsilon e^{-i\alpha_T}\cos \theta ]\nonumber\\
M_2&=&\sqrt{3}m_{3/2}[e^{-i\alpha_S}\sin \theta -
(-1+\delta_{GS})\epsilon e^{-i\alpha_T}\cos \theta ]\nonumber\\
M_1&=&\sqrt{3}m_{3/2}[e^{-i\alpha_S}\sin \theta -
(-\frac{33}{5}+\delta_{GS})\epsilon e^{-i\alpha_T}\cos \theta ],
\end{eqnarray}
in which $\epsilon$, $\epsilon^{'}$ are numerical factors which depend
on the VEV's of $S$ and $T$ (their magnitudes will be discussed below).

We do not consider all possibilities for the $\mu$ and $B$ terms, and
refer the reader to \cite{bim,orbimu} for further discussions of this
issue. We instead first analyze the case in which 
we assume that the $\mu$ problem is solved via an effective coupling in
the superpotential of the form $\mu(S,T)H_1H_2$, (in which $\mu$ depends
only weakly on $S$ and $T$), and then treat $\mu$ and $B$ as independent
parameters. In the first case, the $B$ term is given by
\begin{equation}
\label{bmu1}
B_{\mu}=m_{3/2}[-1-\sqrt{3}e^{-i\alpha_S}\sin \theta -
(1-\frac{\delta_{GS}}{24 \pi^2 Y})^{-1/2}e^{-i\alpha_T} \cos \theta ],
\end{equation}
in which 
\begin{equation}
\label{ydef}
Y=S+S^{*}-\frac{\delta_{GS}}{8\pi^2}\log (T+T^{*}).
\end{equation}
In the $O-II$ scenario described in \cite{bim}, the numerical values of
$\epsilon$ and $\epsilon^{'}$ are taken to be $\sim {\cal O}(10^{-3})$,
which corresponds to the situation in which the VEV's
of ${\rm Re} S$ and ${\rm Re} T$ are ${\cal O}(1)$.  These values are
motivated by the minimization of the scalar potential for $S$ and $T$ that
can be derived either in the gaugino condensation approach
\cite{decarlos,tduality} or more generally imposing the requirements of
$T$-duality on the scalar potential for the modulus field
\cite{tduality,cvetic}.  Within our phenomenological approach,
we can in principle  regard these VEV's as free parameters and
consequently vary $\epsilon$, $\epsilon^{'}$ within reasonable limits;
however, we choose in general not to depart significantly from the case in
which the VEV's are ${\cal O}(1)$.

In addition, a comment about the phases is in order.  While $\epsilon^{'}$
is a real parameter by definition
\begin{equation}
\label{epspr}
\epsilon^{'}=\frac{1}{24\pi^2Y},
\end{equation}
$\epsilon$ is by nature complex \cite{bim}
if the VEV's of $S$ and $T$ are complex, which is of course assumed
throughout this paper to obtain nontrivial values for $\alpha_S$ and
$\alpha_T$. It is also clear
that the phase of $\epsilon$ and $\alpha_{S,T}$ are correlated, with the
particular relations depending on the nature of the nonperturbative
dynamics responsible for the breakdown of supersymmetry.  

In orbifold compactifications within the gaugino condensation approach, in
which the nonperturbative superpotential for $S$ and $T$ takes the form
$W\sim {\rm exp}^{-3f(S,T)/2\beta}$ (with $f(S,T)$ the gauge
kinetic function and $\beta$ the beta-function of the gauge group of the 
gaugino condensate), the soft breaking terms have been computed in
\cite{decarlos} and an analysis of the CP- violating phases has been
carried out explicitly in \cite{bailin1,bailin2}.   The conclusion of
\cite{bailin1,bailin2} is that the properties of the nonperturbative
superpotential (in particular, that the $T$ dependence of the
nonperturbative superpotential $W(T)\sim
\eta(T)^{-6}$) are such that the CP- violating phases of the resulting
soft terms are negligible.  
In their analysis, the VEV's of $S$ and $F^{S}$
are assumed to be real; in principle, the details depend on
the mechanism utilized for the stabilization of the dilaton, which is
usually achieved either through nonperturbative corrections to the
K\"{a}hler potential or through multiple gaugino condensates (``racetrack"
models).  With this assumption and the knowledge of the $T$ dependence of
the superpotential from the form of $f(S,T)$, the value of the phase of
$\epsilon$ can be determined.
The result is that the gaugino masses are
strictly real, as of course are the soft terms which depend solely on
$\alpha_S$ (the issue of the phases of $\mu$ and $B$ is considerably more
complicated and model-dependent).
This result may indicate (as is emphasized  in \cite{bailin1,bailin2})
that in the gaugino condensation approach the phases may be small due to
the properties of the $T$-dependent modular functions.  However,
it was also noted in \cite{cvetic,bailin2} that in principle the
superpotential may depend on other modular invariant functions (such
as the absolute modular invariant $j(T)$) for which the conclusions about 
negligible CP- violating phases may no longer be valid \cite{bailin2}.
We prefer to follow \cite{bim} and not restrict our consideration
to any particular scenario for the supersymmetry breaking, which in turn
allows us to explore the possibility of nontrivial phases for
$\alpha_{S,T}$ and $\epsilon$, which we can treat as independent
parameters in our analysis.   The contrast between the approaches
illustrates the possibility that a measurement of the soft phases will
help determine how supersymmetry is broken and how to relate
compactification to observables.

\subsection{Results}

We start our numerical analysis of the moduli dominated $O-II$ scenario 
by calculating the soft breaking parameters at the electroweak scale.  The
boundary conditions of Eq.~(\ref{soft1} - \ref{bmu1}) are to be
implemented at the string scale $M_{String} \sim 5\times 10^{17}\,
\rm{GeV}$, which in perturbative heterotic string theory is the scale at
which the gauge couplings are predicted to unify \cite{kaplunovsky}, and
all relevant soft breaking parameters are evolved down to the electroweak
scale using two-loop renormalization group equations (RGE's) for the gauge
couplings and one-loop equations for the Yukawa couplings and the soft
parameters\footnote{The discrepancy between the string scale and the GUT
scale $M_{GUT}\sim 2\times 10^{16}\,\rm{GeV}$ (where the gauge couplings
appear to unify from extrapolating the measured values of the electroweak
scale couplings to higher scales assuming the MSSM particle content) is a
well-known problem in perturbative heterotic string theory with a number
of solutions proposed (see e.g. \cite{dienes}).  In practice, this
mismatch  between the scales introduces a small numerical discrepancy
into the analysis (unless as in \cite{bim} intermediate scale matter or
some other effect is assumed to be present which solves the problem of the
unification of the couplings).}.  

The renormalization group analysis and resulting patterns for the
low-energy mass spectrum of the soft terms of the $O-II$ scenario 
(assuming $\alpha_{S,T}=0$) have been presented in \cite{bim,gunion,casas},
to which we refer the reader for further details. As noted in the previous
section, the gaugino mass phases do not run at one-loop; this behavior
can be disrupted only by higher loop corrections, threshold 
effects and other possible corrections \cite{graham}. However, the
trilinear coupling  phases $\varphi_{A_u}$, $\varphi_{A_d}$,
$\varphi_{A_t}$ and $\varphi_{A_e}$ do run, and they evolve away from the
single universal string scale value at different rates depending on the
relevant Yukawa couplings and gauge group charges. 
We perform an $R$ rotation at the electroweak scale, which allows us to 
set the phase of $M_2$ equal to zero.  The phase of $\mu$ is then
determined by the phase of the $B$ parameter as 
$\varphi_{\mu}=-\varphi_{B}$, so that $B\mu$ is real and the Higgs
potential is not affected by the phases at tree level.  

As previously mentioned, we regard $m_{3/2}$, $\delta_{GS}$, $\theta$,
$\epsilon$, $\alpha_S$, and $\alpha_T$ as the free parameters of the
model.
Since  $\epsilon$ is $\sim O(10^{-3})$ (assuming the VEV's of $S$ and $T$ 
are ${\cal O}(1)$), it is necessary to consider the small $\theta$
(moduli-dominated) limit for the gauginos to acquire significant relative
phases.   We consider a range of numerical values of
$\theta$ between $10^{-3}$ and $10^{-1}$, which in turn requires $m_{3/2}$
to be typically greater than $O(\rm{TeV})$ for the soft breaking
parameters to have acceptable masses.
In addition, if the $B$ term condition (\ref{bmu1}) is imposed, such a
large mass scale for the gravitino could cause the $B_{\mu}$ parameter to
be of the same order, which is disfavored by naturalness arguments.  One
way to avoid this result is to require $\alpha_T\sim\pi$ for
$\cos\theta\simeq 1$ or
$\alpha_T\sim 0$ for $\cos\theta\simeq -1$.
It is also clear that if $\delta_{GS}>>5$ or so,  the relative gaugino
phases will be suppressed producing no interesting CP- violating
phenomena.  To obtain a large relative phase between gaugino masses $M_i$
and  $M_j$, it is necessary that 
\begin{equation}
\label{ineq}
\min (\kappa_i,\kappa_j) |\epsilon | \alt |\theta  | \alt 
\max (\kappa_i,\kappa_j) | \epsilon  | ,
\end{equation}
where $\kappa_1 =(-\frac{33}{5}+\delta_{GS})$, $\kappa_2=(-1+\delta_{GS})$ and
$\kappa_3=(3+\delta_{GS})$. All of the above relations significantly
constrain the possible parameter space and introduce further correlations
between the phases.  In particular, the seven CP-violating  phases
entering the calculation of the electric dipole moments are effectively
parametrized by a single phase $\alpha_S$ originating in the string
sector\footnote{Despite
different approaches, our results are in a sense consistent with the
results of \cite{bailin1,bailin2} in the gaugino condensation framework,
in which they obtain negligible phases in the soft breaking terms (in
particular in the gaugino mass sector) with the assumption of vanishing
phases for the scalar and $F$- component VEV's of the dilaton.}.
Despite this high degree of correlation, we find that it is
possible to find parametric configurations that lead to relatively
substantial values of the phases and light superpartner mass spectra while
satisfying the electric dipole moment constraints.

We explore the possibility of obtaining superpartner mass
spectra with the squark masses lying below $500\, \rm GeV$ while
simultaneously generating sizeable and experimentally acceptable CP-
violating phases at the electroweak scale within the $O-II$ string
scenario for a series of parameter sets classified by the value of the
Green-Schwarz parameter $\delta_{GS}$.
In addition to the general $O-II$ relations
(\ref{soft1})-(\ref{gaugmass1}), we consider two general cases.  First, we
impose the $B$- term condition (\ref{bmu1}), corresponding to a particular
solution to the $\mu$ problem (arising from nonperturbative corrections to
the superpotential). To consider the case in which $\mu$ and $B$ can
receive contributions from other sources, such as from the Giudice-Masiero
mechanism, we do not discuss all of the possibilities in detail,
but rather relax the expression  (\ref{bmu1}) for $B$ and treat $B$ as a
free complex parameter. This approach adds one more CP phase, namely
$\varphi_{\mu}$, to the set of independent parameters. As in the previous
case, $\mu$ is still undetermined and hence is regarded as a free
parameter.  In addition, the restriction on $\alpha_T$, which was
dictated by the form of (\ref{bmu1}), can in principle be
relaxed.  However, it is important to note that the expression for the
$B$ term arising in the Giudice-Masiero mechanism suffers from a similar
problem \cite{muproblem}, and thus this restriction is likely to be
generic (barring possible cancellations in the $B$ term arising from
different sources of the $\mu$ term, which is a possibility we do not
consider further in this paper).  
We find in general that the differences between the case in which the $B$
term is determined within the string model through (\ref{bmu1}) and the
case in which $B$ is left as an independent parameter are not very
significant.  Therefore, in the results which follow, we display the
results for the two cases in tandem to emphasize this feature.

We can further determine several general constraints on the parameter
space of this model. As the scalar masses are universal at the string
scale, the color-neutral scalar particles ({\it i.e.}, sleptons and
sneutrinos) are significantly lighter than the
squarks due to smaller RGE running, with masses typically $\simlt 200\,
\rm GeV$. The choice of $\delta_{GS}$ and the requirement of light
sfermion masses effectively determines the range of values for the
gravitino mass parameter $m_{3/2}$ from (\ref{soft1}), and  
we take $\epsilon^{'}\sim 10^{-3}$ (following \cite{bim}).
We can also estimate the interesting range  of $\theta$ and $\epsilon$
providing for light gauginos. To obtain optimally
large $\varphi_1$ and $\varphi_3$ we set $|\theta|\simeq|\kappa_2
\epsilon|$. The values of the universal $A$ parameters (and the $B$ term
when (\ref{bmu1}) is imposed at the string scale) are then fully
determined by the choice of $\delta_{GS}$, $m_{3/2}$, $\theta$,
$\epsilon$,
and $\alpha_S$ \footnote{We note that the RGE running of $\varphi_{A_i}$
is the same as in supergravity models, and has been discussed for instance
in \cite{sugracp}.}. 

\begin{figure}[h!]
\centering
\epsfxsize=6in
\hspace*{0in} 
\epsffile{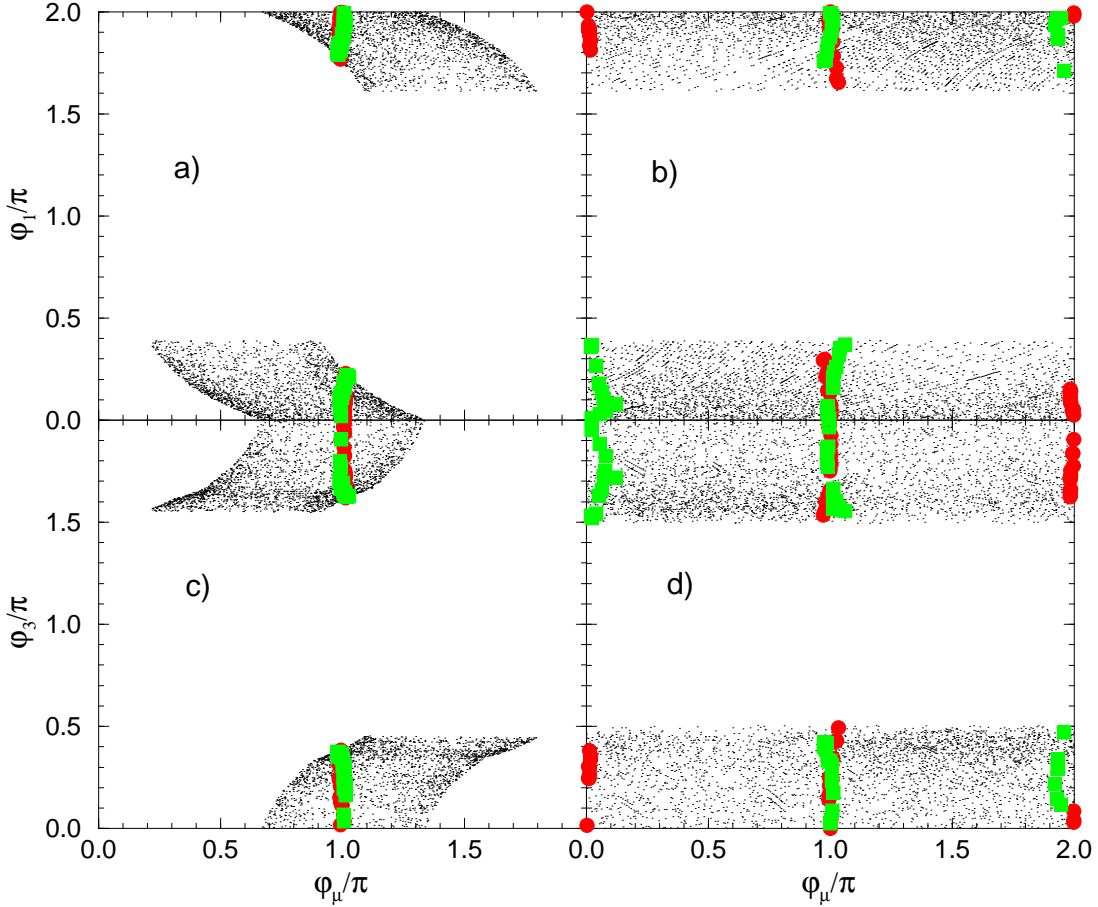}
\bigskip
\caption{Regions allowed by the electron and neutron EDM constraints in the 
$O-II$ scenario for $m_{3/2}=4\, \rm{TeV}$, $\delta_{GS}=-2$, $\theta=0.021$
$\epsilon=0.007$, $\tan\beta =2$ and $\mu=100\, \rm{GeV}$. 
The dotted areas show allowed regions resulting from the specific form
of soft breaking parameters. The red (black)
circles denote points allowed by the eEDM and the green (grey) blocks by the
nEDM. In frames  $a)$ and $c)$, the $B$ term is assumed to originate from 
an effective coupling in the superpotential (Eq.~\ref{bmu1}) while in
frames  $b)$ and $d)$ $B$ is  treated as an independent parameter and 
its magnitude was set to  $|B|=300\, \rm{GeV}$.  
} 
\label{figone} 
\end{figure}

In Figure \ref{figone}, we plot the regions of the three most important
phases 
$\varphi_{\mu}$,  $\varphi_{1}$, and $\varphi_{3}$ for the case of
$\delta_{GS}=-2$.
Frame $a)$ shows the points allowed by the electron and neutron EDM
constraints in the $\varphi_{\mu}-\varphi_{1}$ plane, while frame $c)$
delineates the projection of these points onto the
$\varphi_{\mu}-\varphi_{3}$ plane.  In frames $b)$ and $d)$, we display
the results for the same sets of parameters but taking $B$ as an
additional independent parameter which we set to the value $|B|=300\,\rm 
GeV$.

The results illustrate a general feature of this model:
to obtain an overlap between the neutron EDM allowed and the
electron EDM allowed regions in this model, a low value of $\mu$ is
required. This restriction implies large gaugino-higgsino mixing, and thus
rather small values of  $\varphi_{\mu}$ are needed to satisfy the electron
EDM constraint, as discussed in \cite{bgk}. In particular, if $\mu$ is
increased the electron and neutron regions overlap only in the small phase
region where all phases are $\simlt 10^{-2}$.  Small values of
$\varphi_{\mu}$ illustrate that the chargino contribution to the electron
EDM is generally much larger than the corresponding neutralino
contribution, and hence the cancellation between these contributions  is 
not adequate. As a result, both contributions have to be suppressed by
small
values of $\varphi_{\mu}$;  residual cancellation in the vicinity of 
$\varphi_{\mu}\sim \pi$ subsequently ensures that the effects of $\varphi_{1}$ 
and $\varphi_{A_e}$ also cancel. A similar effect takes place
in the case of the neutron dipole moment, with similar restrictions on the
corresponding relevant phases. The overlap between the two regions then
yields $\varphi_{1}\sim \frac{\pi}{6}$ and remarkably, $\varphi_{3}\sim
\frac{\pi}{2}$.

\begin{figure}[h!]
\centering
\epsfxsize=6in
\hspace*{0in} 
\epsffile{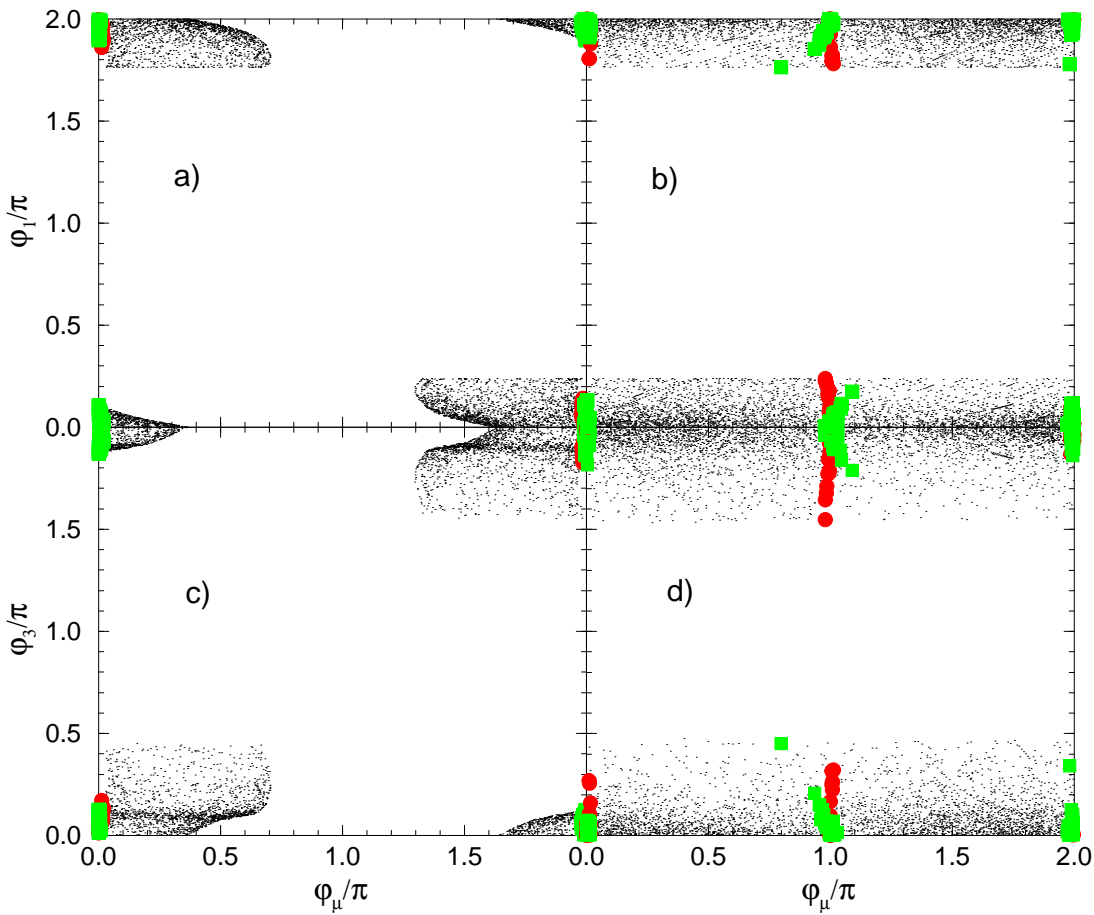}
\bigskip
\caption{Same as Fig.~\ref{figone} but for $m_{3/2}=2\, \rm{TeV}$, 
$\delta_{GS}=-10$, $\theta=0.06$ and  $\epsilon=0.006$. } 
\label{figfour} 
\end{figure}

General arguments show that for $|\delta_{GS}|\simlt 5$ and
$O(1)$ $\varphi_{1}$ and $\varphi_{3}$, viable solutions can be
obtained provided $\varphi_{\mu}$ is close to $\pi$ (or zero).
However, the accessible values of
$\varphi_{\mu}$, $\varphi_{1}$ and $\varphi_{3}$ are reduced
as $|\delta_{GS}|$ increases,  and
hence it is more difficult to satisfy the EDM constraints with
large phases.  Figure \ref{figfour} clearly demonstrates this effect for
$\delta_{GS}=-10$; in this case the gaugino masses are approximately
universal, and correspondingly the phases of soft breaking parameters are
constrained to satisfy the traditional bound.
The results demonstrate that the range of values of $\delta_{GS}$
for which the phases of the soft terms are nontrivial is quite
restricted (but are within a reasonable range of values determined in
explicit orbifold models).  Therefore, if the model of this section were
the way nature behaved, it would be possible to determine the anomaly
cancellation parameter $\delta_{GS}$ by the measurements of the EDM's.

We close this section with a brief comment about a further generalization
of the $O-II$ scenario, in which the assumption that only the single
modulus $T$ plays a role in supersymmetry breaking is
relaxed. For the purposes 
of this study, the important feature remains that these individual
moduli will contribute to gaugino masses only at one-loop.  Therefore,
the number of parameters increases substantially; in addition to the 
need to define extra Goldstino angles (as is done in \cite{bims} and will
be required in the Type I models discussed below), there will be
\cite{bimrev,orbifolds} a $\delta_{GS}$, $\epsilon$, etc. for each modulus
field involved in the supersymmetry breaking, the details of which will
depend on the particular orbifold model under consideration.  Due to the
additional complications
and model-dependence, we do not consider such scenarios further in this
paper. We anticipate that in general there can be particular models for
which the parameter space for viable large phase solutions will be wider
than that of the minimal scenario considered in this paper. 

\section{Soft Breaking Terms in Ho\v{r}ava-Witten  Scenarios}
\label{IV}
\subsection{Theoretical Framework}

We now turn to a newer class of models based on the work  of Ho\v{r}ava
and Witten \cite{horavawitten}, who showed that eleven-dimensional
supergravity (the
conjectured low energy limit of M theory) compactified on a Calabi-Yau
three-fold times an orbifold interval along the eleventh dimension gives
rise to $E_8 \times E_8$ gauge theories with $N=1$ supersymmetry in four
dimensions, and further proposed that this framework describes the
strongly coupled heterotic $E_8 \times E_8$ string theory.  In this
scenario, the two $E_8$ gauge multiplets reside on two ten-dimensional
boundaries, which are separated by the interval corresponding to the
eleventh dimension.  The phenomenological implications of this scenario
display several attractive features which are not present in the case of
perturbative heterotic string theory.  For example, there is the 
possibility of reconciling the string scale and the GUT scale, which is an
encouraging result for the unification of the gauge couplings.
Furthermore, the usual hidden sector mechanism for the breakdown of
supersymmetry can be naturally realized in this class of
models; supersymmetry can be broken (perhaps via gaugino condensation) on
the hidden boundary, and transmitted to the observable sector by the
dilaton and moduli fields, which can travel in the bulk (see 
e.g. \cite{nilles,munoz}).  

In \cite{munoz}, the soft supersymmetry breaking parameters were derived
within the framework in which the effects of supersymmetry breaking are
encoded in the parameterization (\ref{fvevs0}) of the auxiliary component
VEV's of the dilaton $S$ and overall modulus $T$.  The results were
obtained by determining the form of the K\"{a}hler potential,
superpotential, and gauge kinetic function to the first subleading order
in the M theory expansion of the effective four-dimensional supergravity
theory\footnote{It is important to note that the  many studies of M
theory vacua assume the ``standard embedding", in which the spin
connection is embedded into one of the $E_8$ gauge groups, although 
the standard embedding does not play a
special role in the construction of these vacua (in contrast to the case
of the weakly coupled heterotic string) \cite{ovrut}. While in fact the
relaxation of this condition can lead to more general scenarios, the
conclusions about nontrivial CP effects are the same for the case in which
$S$ and a single $T$ modulus contribute to the supersymmetry breaking.}. 
For the purposes of this study, we note that the gauge kinetic
function of the observable sector $E_8$ gauge group takes the form
\begin{equation}
\label{fmth}
f_{obs}=S+\alpha T,
\end{equation}
in which $\alpha$ is a coefficient of ${\cal O}(1)$.  This feature is in
direct contrast to the $T$- dependent piece of the gauge kinetic function
in the perturbative case (\ref{f1loop}), which is suppressed by a
loop factor.  The soft breaking parameters take the form
\begin{eqnarray}
\label{softm}
M&=&\frac{\sqrt{3}m_{3/2}}{1+\epsilon_0}\left (\sin \theta
e^{-i\alpha_S}+\frac{\epsilon_0}{\sqrt{3}}\cos \theta e^{-i\alpha_T}
\right )\nonumber\\
m^2&=&m^2_{3/2}-\frac{3
m^2_{3/2}}{(3+\epsilon_0)^2}(\epsilon_0(6+\epsilon_0)\sin^2 \theta
+(3+2\epsilon_0)\cos^2 \theta \nonumber\\&-&2\sqrt{3}\epsilon_0 \sin
\theta \cos \theta
\cos (\alpha_S-\alpha_T))\nonumber\\
A&=&-\frac{\sqrt{3}m_{3/2}}{3+\epsilon_0}\left ((3-2\epsilon_0) \sin
\theta e^{-i\alpha_S}+\sqrt{3}\epsilon_0 \cos \theta e^{-i\alpha_T}
\right ),
\end{eqnarray}
in which $\epsilon_0$ is given by 
\begin{equation}
\epsilon_0=\frac{4-(S+S^*)}{S+S^*}.
\end{equation}
As discussed in \cite{munoz} (to which we refer the reader
for an explanation of this point), the  standard embedding constrains the
range of $\epsilon_0$ to  $0 < \epsilon_0 <1$.  

It is clear from above relations that in these scenarios, the gaugino mass
parameters are universal when consideration is restricted to the case in
which the dilaton and the single $T$ modulus participate in supersymmetry
breaking.  Therefore, we anticipate that the cancellation mechanism
generally will not be adequate, from the general discussion presented in
Section \ref{II}.   We note that this conclusion is not likely to hold in
more general scenarios in which several individual $T_m$ moduli associated
with the Calabi-Yau manifold are involved in the supersymmetry breaking.
In the multimoduli case, it is likely that the $T_m$-dependent
contributions to the gaugino masses will be gauge-group dependent; if
these contributions have ${\cal O}(1)$ coefficients as in (\ref{fmth}),
the gaugino masses will be nonuniversal over a greater range of parameter
space than in the $O-II$ orbifold model discussed in the previous section,
and may provide for interesting CP effects. However, we restrict our
consideration to the overall modulus case in this paper, and defer the
study of more generalized Ho\v{r}ava-Witten scenarios to a future study.

\subsection{Results}

In our analysis of possible CP effects in the Ho\v{r}ava-Witten scenario
we proceed along similar lines as in the  $O-II$ orbifold models.
However, in principle there is an important difference in that the string
scale is not fixed to the value $M_{String} \sim 5 \times 10^{17}\,
\rm{GeV}$ as in the heterotic case, but can take any value
\cite{horavawitten},  including $M_{G}$ (which we choose for
simplicity).
We start from the free parameters $m_{3/2}$, $\theta$, and $\epsilon_0$, 
which, in combination with the two independent complex phases $\alpha_S$
and $\alpha_T$, determine the soft breaking parameters at the string
scale. All soft terms are subsequently  RGE evolved from  $M_{String}$
down to the electroweak scale and particle masses are calculated together
with all physical CP- violating phases. We consider $B$ and $\mu$ to be
independent parameters, although their magnitudes are numerically
determined from the
requirement of radiative symmetry breaking. The phase of $\mu$ is varied 
independently as another free parameter of the model.   

\begin{figure}[h!]
\centering
\epsfxsize=4in
\hspace*{0in} 
\epsffile{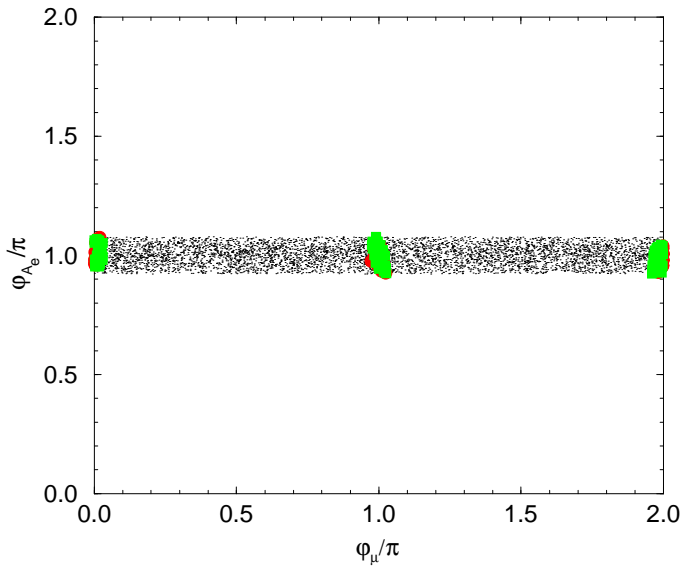}
\bigskip
\caption{Electron and neutron EDM allowed regions for the
Ho\v{r}ava-Witten  
scenario with $m_{3/2}=500$ GeV, $\theta=0.5$, $\epsilon_0=0.9$, 
and $\tan\beta=2$ shown in the $\varphi_{\mu}$-$\varphi_{A_e}$ plane. 
The dotted area shows the region of phases allowed in this scenario.
The red (black) circles and green (grey) blocks denote points allowed 
by the eEDM and nEDM respectively. 
The values of $|B|$ and $|\mu|$ are fixed by radiative electroweak 
symmetry breaking. 
} 
\label{figfive} 
\end{figure}

The relative phase  between the universal values of $M$ and $A$ determines
the physical CP- violating phases of the soft $A$ parameters at the string
scale. It is obvious that its value is restricted by the allowed ranges of 
$\epsilon_0$ and $\theta$; for example, in the limiting cases when either
$\sin\theta$ or $\cos\theta$ are zero, the relative phase is zero at the
string scale.
The additional requirement of positive $m^2$ values also restricts the
allowed regions of $\epsilon_0$ and $\theta$. As a result, it is 
difficult to obtain large phases of the $A$ parameters at the electroweak
scale as  $\alpha_S$ and $\alpha_T$ are varied from zero to $2\pi$.
In Figure \ref{figfive} we show the points allowed by the electron and
neutron EDM's in
a typical example of this scenario with $m_{3/2}=500\, \rm{GeV}$,
$\epsilon_0=0.9$, and $\theta=0.5$. Since the gaugino phases are
identically zero and the range of $\varphi_{A_e}$ is severely restricted
by correlations between $A$ and $M$ at the string scale, the
cancellations are insufficient in this particular scenario.   
Therefore, only a very small fraction of the
$\varphi_{\mu}-\varphi_{A_e}$ parameter
space leads to models allowed by the electron EDM, in direct analogy
with the dilaton-dominated scenario in the perturbative heterotic models
discussed in \cite{barrkhalil}.

\section{Soft Breaking Terms in Type I Models}
\label{V}
\subsection{Theoretical Framework}

We now turn to another example of a new class of models,
the four-dimensional Type IIB
orientifold models with $N=1$ supersymmetry
\cite{sagnotti,kakushadze,ibanez,shiu}. These
models are based on the Type IIB (closed string) theory compactified on
orientifolds, which are orbifold compactifications accompanied by an
additional worldsheet parity operation.  The consistency of the theory 
requires the addition of open string (Type I) sectors, with the open
strings ending on Dirichlet D- branes. It is important to note that
orientifolds are illustrative
of a much larger class of models in the Type I picture, containing
more general configurations of nonperturbative objects (e.g. D- brane
bound states) in more general singular backgrounds (e.g. conifolds
\cite{conif}).

The number and type of D- branes required in a given model depends on
the details of the orientifold group; however, in the most general case
for compact Abelian orbifolds there is one set of nine-branes  and three
sets of five-branes ($5_i$),
in which $i$ labels the complex coordinate of the internal space
included in the five-brane world-volume. Gauge groups are associated with
each set of coincident D- branes.  
These models are constructed utilizing perturbative techniques. However,
due to the Type I-heterotic $S$- duality, the Type IIB orientifold models
have heterotic duals; the heterotic duals are perturbative for orientifold
models with only nine-branes (such as the $Z_3$ orientifold
\cite{sagnotti}), but nonperturbative in the more general case with
additional sets of five-branes (such as orientifolds with 
order-two twists \cite{kakushadze}).

The chiral matter fields also arise
from open string sectors and can be classified into two categories.  The
first category are fields which arise from open strings which start and
end on the same type of D- branes. These fields are therefore charged
under only the (generically non-Abelian) gauge group of that set of
branes, typically in the fundamental or antisymmetric tensor
representations. The second class of fields originate from open strings
which start and end on different types of branes and are hence are charged
under the two associated gauge groups.  In this case, the states are
bifundamental representations under the associated two gauge groups from
the two D- brane sectors.  In the closed string sector, there are the 
dilaton $S$ and moduli fields $T_i$, as well as the twisted
sector moduli, which play a role in the cancellation of the anomalous
$U(1)$'s generically present in these models \cite{typeIanom}.

A recent investigation shows that the phenomenological
properties (including the possibilities for gauge coupling unification)
\cite{ibanez} of these models are quite distinctive from those of
the perturbative heterotic models.  In particular, the string scale
is not fixed close to $M_{Planck}$ as in the weakly coupled heterotic
case,  but rather can take lower values. The implications for electroweak
scale physics also crucially depend on the nature of the embedding of the
SM gauge group into the different
D- brane sectors.

The soft supersymmetry breaking terms obtained when the dilaton and moduli
fields are responsible for the breakdown of  supersymmetry can be
determined using the  parameterization of 
the $F$-component VEV's (following \cite{bims,ibanez}):
\begin{eqnarray}
F^S&=&\sqrt{3}(S+S^{*})m_{3/2}\sin \theta e^{i\alpha_S}\nonumber\\
F^i&=&\sqrt{3}(T_i+T^{*}_i)m_{3/2} \cos \theta \Theta_i e^{i\alpha_i},
\end{eqnarray}
in which $\Theta_i$ are generalized Goldstino angles (with $\sum_i
\Theta_i^2=1$).  The soft terms can then be computed \cite{ibanez} with
the knowledge of the structure of the Yukawa superpotential couplings
\cite{sagnotti,kakushadze} and the tree-level K\"{a}hler potential and
gauge kinetic functions \cite{ibanez}, which have also been determined for
this class of models.

For the purposes of the studying the phase structure of the soft terms, we 
note that the gauge kinetic functions determined in \cite{ibanez} take the
form
\begin{eqnarray}
\label{orientifoldf}
f_{9}&=&S\nonumber\\
f_{5_i}&=&T_i,
\end{eqnarray}
which demonstrate that the dilaton no longer plays a universal role (as
the moduli dependence now occurs at the tree-level) as it did in the
perturbative heterotic case.  In particular, the structure of
(\ref{orientifoldf}) illustrates a distinctive feature of this class of
models, which is in a sense there is a different ``dilaton" for each
type of brane.

This fact has important implications in this class of models both for
gauge coupling unification \cite{ibanez} and the  patterns of gaugino
masses, which strongly depend on the embedding of the SM gauge group.
In the case in which the SM gauge group is embedded in a single D- brane
sector, the pattern of the gaugino masses resembles that of the tree-level
gaugino masses in the weakly coupled heterotic models studied in the
previous section.  For example, if the SM gauge group is embedded within
the nine-brane sector, this can be seen from the similarity between
(\ref{orientifoldf}) and the corresponding tree-level expression for $f$
in the perturbative heterotic models (\ref{ftree}); 
the situation is similar (with the corresponding modulus field $T_i$
playing the role of the dilaton) if the SM arises from a single $5_i$
brane sector.

However,  if the SM gauge groups arise from distinct D- brane sectors,
there is the possibility of nonuniversal gaugino masses at the tree-level,
which can be seen from (\ref{orientifoldf}).  This feature was not
possible in the perturbative heterotic models discussed in the previous
sections, and is interesting from the point of view of obtaining
new patterns of nontrivial relative phases in the soft terms. 

To explore this possibility, we consider toy models of soft terms derived
with the assumption that the $SU(3)$ and $SU(2)$ gauge groups arise from
different five-brane sectors\footnote{We could also assume that one of the
gauge groups arises from the nine-brane sector.  It was noted in
\cite{ibanez} that it may be more difficult to obtain consistent
unification of the gauge couplings at the GUT scale in this case.
Although this point is not crucial for the purposes of this study, we
choose the case of embedding the SM in the five-brane sectors for the sake
of definiteness.} (for example, $5_1$ and $5_2$).  The possibilities for
the embedding of $U(1)_Y$ are then restricted by phenomenological
criteria. For example, an important constraint is that the MSSM particle
content contains the quark doublet states, which are charged under all of
the gauge groups; this fact restricts $U(1)_Y$ to arise from the $5_1$
and/or $5_2$ sectors, as the matter fields of these Type I models are at
most charged under the gauge groups of {\it two} D- brane sectors.  
In this paper, we choose for simplicity to restrict our consideration to
simplified scenarios in which $U(1)_Y$ resides in either the $5_1$ or the
$5_2$ sector \footnote{Although it is not clear if such special cases can 
be realized in explicit orientifold models, we note that in the models
that have been constructed to date in which the SM non-Abelian gauge
groups can arise from different D-brane sectors, the hypercharge gauge
group is in general a linear combination of gauge groups arising from the
two sectors. We thank Gary Shiu for a discussion of this point.}.
Depending on the details of the hypercharge embedding,
the remaining MSSM states may either be states which (in analogy with the
quark doublets) are trapped on the intersection of these two sets of
branes, or states associated with the single $5_i$ sector which contains
$U(1)_Y$. In any event the natural starting point for constructing models
with these features are orientifolds which realize identical GUT gauge
groups and massless matter on two sets of intersecting 5- branes.
The existence of such symmetrical arrangements is often guaranteed by
T-duality. For example, Shiu and Tye \cite{shiu}
have exhibited an explicit model which realizes the Pati-Salam gauge
fields of $SU(4)$$\times$$SU(2)_L$$\times$$SU(2)_R$ and identical chiral
matter content on two sets of 5- branes.
Additional Higgsing and modding by discrete symmetries could then in
principle produce the asymmetrical structures
outlined above.

In this scenario, the soft scalar masses can take the form
(see the general formulae in \cite{ibanez}):
\begin{eqnarray}
\label{model3}
m^2_{5_15_2}&=&m^2_{3/2}(1-\frac{3}{2}(\sin^2 \theta+\cos^2 \theta
\Theta_3^2) ) \nonumber\\
m^2_{5_1}&=&m^2_{3/2}(1-3\sin^2 \theta ).
\end{eqnarray}
In the case with $U(1)_Y$ and $SU(3)$ from the $5_1$ sector,
the $SU(2)$ doublet states clearly arise from open strings stretching
between the two D- brane sectors, while the $SU(2)$ singlets can either
be states of the same type or states associated with the $5_1$ brane
sector only.   The gaugino masses and $A$ terms take the form 
\begin{eqnarray}
\label{model4}
M_{1}&=&\sqrt{3}m_{3/2}\cos \theta \Theta_1
e^{-i\alpha_1}=M_3=-A_{t,e,u,d}\nonumber\\
M_{2}&=&\sqrt{3}m_{3/2}\cos \theta \Theta_2 e^{-i\alpha_2},
\end{eqnarray}
and hence the relations among the phases $\varphi_i$ of the gaugino mass
parameters $M_i$ are $\varphi_1=\varphi_3\neq \varphi_2$.
Similar expressions apply for the case in which $U(1)_Y$ and $SU(2)$ are
associated with the same five-brane sector; in this case, the relations
among the phases $\varphi$ of the gaugino mass parameters are 
$\varphi_1=\varphi_2\neq \varphi_3$.
In these models, the solution to the $\mu$ problem is not certain, and
hence $\mu$ and $B$ are free complex parameters in the analysis (although
their phases are as usual related by the PQ symmetry of the MSSM
superpotential).  

Due to the absence of quasi-realistic Type I models as
yet (despite continued progress in model-building techniques
\cite{sagnotti,kakushadze,shiu}), it is not certain whether this type of
SM embedding can be realized in an explicit orientifold model.
Therefore, we emphasize that these models should be
interpreted as toy models which illustrate new possibilities for the
patterns of soft breaking terms in this new class of four-dimensional
superstring models.

\subsection{Results}

Our numerical analysis of the Type I models closely follows
the approach adopted for the Ho\v{r}ava-Witten scenarios.  In the Type I
models, the string scale is not fixed;
for the sake of simplicity, we assume the
string scale and the GUT scale coincide, and that the gauge
couplings unify at this scale. It is beyond
the scope of this paper to consider all
possibilities, and we refer the reader to a comprehensive discussion of
this issue and its implications for gauge coupling unification in
\cite{ibanez}. 
For the sake of simplicity, we assume the
string scale and the GUT scale coincide, and that the gauge
couplings unify at this scale.  Thus, in the first model we consider
with $SU(3)$ and $U(1)_Y$ arising from the same brane sector, the boundary
conditions (\ref{model3}) and (\ref{model4}) are implemented at the 
GUT scale, and the parameters are subsequently evolved down to the
electroweak scale. The sparticle masses and the CP-violating phases
depend on the free parameters $m_{3/2}$, $\theta$, $\Theta_i,\ i=1,2,3$,
which are related by $\Theta_1^2+\Theta_2^2+\Theta_3^2=1$, as well as the
two phases $\alpha_1$ and 
$\alpha_2$. To avoid negative scalar mass-squares we restrict our
consideration to values of $\theta$ which satisfy $\sin^2\theta
<\frac{1}{3}$, and also assume that $\Theta_3=0$ (indicating that the
modulus $T_3$ associated with the $5_3$ brane sector plays no role in
supersymmetry breaking, and thus is essentially decoupled from the
observable sector).
We also treat $B$ and $\mu$ as free parameters, as they are not determined
in this scenario.  In addition, we explore the phenomenologically
motivated scenario in which the electroweak symmetry is broken radiatively
as a result of RGE evolution of the Higgs masses $m_{H_1}^2$ and
$m_{H_2}^2$.  As the minimization conditions are imposed at 
the electroweak scale, the values of $B\mu$ and $|\mu |^2$ can be
expressed in  terms of $\tan\beta$ and $M_Z$ \cite{rad}. However, even
under these assumptions $\varphi_{\mu}$ is still an independent parameter.

\begin{figure}[h!]
\centering
\epsfxsize=6in
\hspace*{0in} 
\epsffile{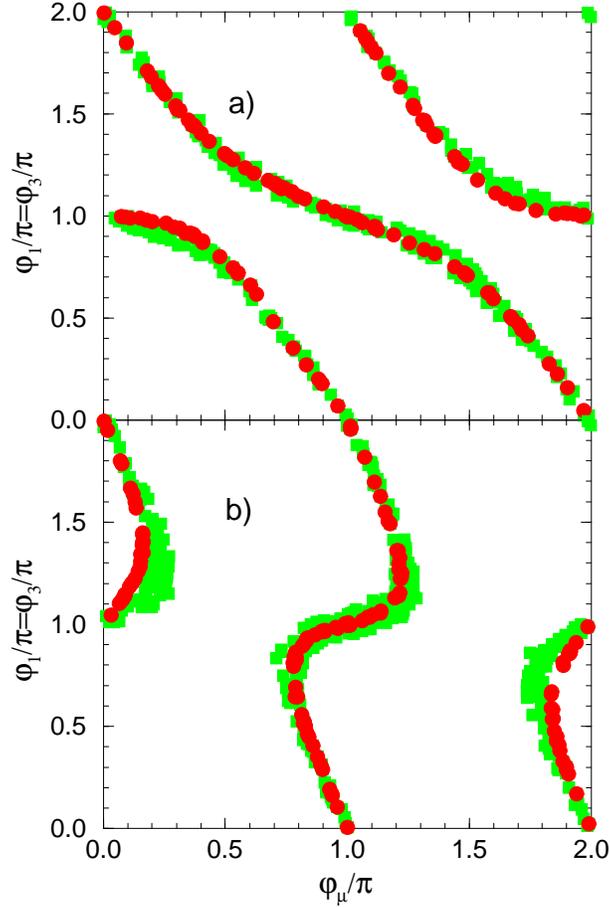}
\bigskip
\caption{Electron (red (black) circles) and neutron (green (grey)
blocks) EDM allowed regions for the Type I
orientifold models with $m_{3/2}=150\, \rm{GeV}$, $\theta=0.4$, 
$\Theta_1=0.85$ and $\tan\beta=2$. In frame $a)$ the values of $B$ and
$\mu$ are assumed to be independent and their magnitudes  are set to 
$|B|=100\, \rm{GeV}$ and $|\mu|=600\, \rm{GeV}$. Frame $b)$ shows
the results for the case when electroweak symmetry is assumed to be
broken radiatively.  
 } 
\label{figsix} 
\end{figure}

In frame $a)$ of Figure \ref{figsix}, we show the results for
$m_{3/2}=150\, \rm
GeV$, $\theta=0.4$
and $\Theta_1=0.85$. As in the previous case of orbifold models we fix 
$\tan\beta=2$, although different values of $\tan\beta$ have been
explored~\footnote{We do not consider large values of $\tan\beta$, as new 
types of contributions can become important \cite{pilaftsis}.}. In this
case we do not impose the condition
of correct radiative electroweak symmetry breaking, but rather assume $B$
and $\mu$ take the values $|B|=100\, \rm GeV$ and $|\mu|=600\, \rm GeV$.
We find here, remarkably, that in order to satisfy the experimental
constraints on the electron and neutron
EDM's in this model, the large individual contributions from chargino, 
neutralino and gluino loops do not have to be suppressed by small CP phases.
A cancellation between the chargino and neutralino loop
contributions thus causes the electron EDM to be acceptably small. As
emphasized in \cite{bgk}, the contributions to chargino and neutralino
diagrams from gaugino-higgsino mixing naturally have opposite signs and
the additional $\varphi_1$ dependence of the gaugino exchange
contribution to the neutralino diagram can provide
for a match in size between the chargino and neutralino contributions.
The importance of the gaugino exchange diagrams increases for large values
of $\mu$ and allows the cancellation to take place for a wider range of
$\varphi_{\mu}$ values. In the neutron case, the contribution of the
chargino loop is offset by the gluino loop contributions to the electric
dipole operator $O_1$ and the chromoelectric dipole operator $O_2$. Since
$\varphi_1=\varphi_3$ in this scenario, the gluino contribution
automatically has the correct sign to balance the chargino contribution
in the same region of gaugino phases which ensures cancellation in the
electron case. This simple and  effective mechanism therefore provides
extensive regions of parameter
space where the electron and neutron EDM constraints are satisfied
simultaneously while allowing for $O(1)$ CP-violating phases.

If electroweak symmetry is assumed to be broken radiatively, the resulting 
value of $|\mu|$ is somewhat smaller. In our particular case with the
remaining parameters unchanged $|\mu|\sim 350\, \rm GeV$.
The ranges of allowed CP-violating phases are shown in Figure
\ref{figsix} $b)$. Here also
the electron and neutron EDM allowed regions overlap substantially although
the range of $\varphi_{\mu}$ is slightly reduced.       
However, the general picture is valid, and low energy models with
light superpartner mass spectra and large CP-violating phases can be
obtained within this framework, even including the constraint of
electroweak symmetry breaking.

\begin{figure}[h!]
\centering
\epsfxsize=6in
\hspace*{0in} 
\epsffile{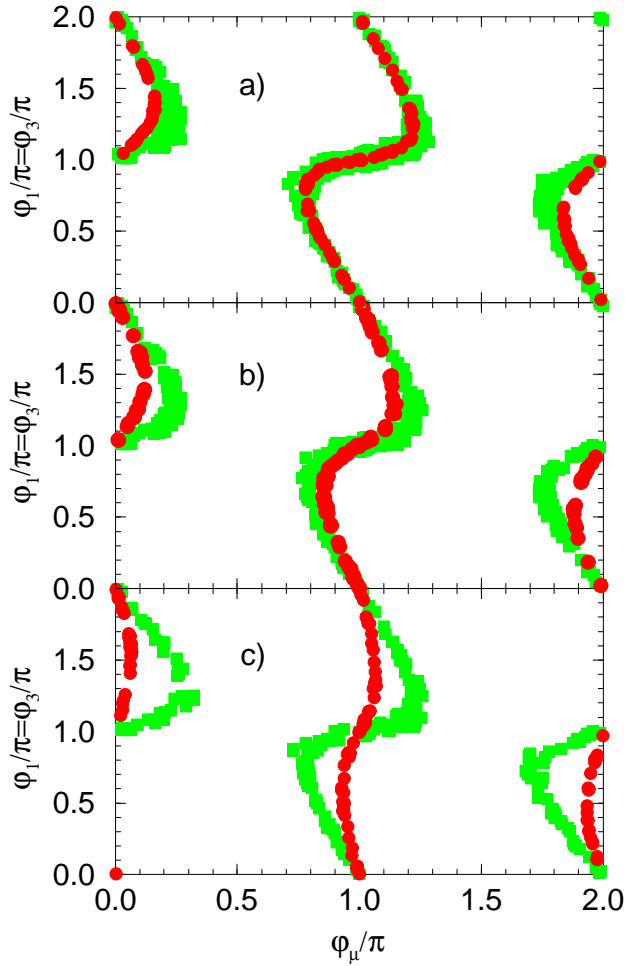}
\bigskip
\caption{Illustration of the overlap between the regions allowed by the
electron EDM (denoted by the red (black) circles) and neutron EDM
(denoted by the green (grey) blocks) constraints. We choose $m_{3/2}=150\,
\rm{GeV}$, $\theta=0.4$ and $\tan\beta=2$, and impose radiative EW
symmetry breaking. Allowed points are shown for {\it
a)} $\Theta_1=0.85$,  {\it b)} $\Theta_1=\sqrt{1/2}$, and {\it c)} 
$\Theta_1=0.55$.} 
\label{figsixbar} 
\end{figure}

The cancellation mechanism in this scenario provides a large
range of allowed CP- violating soft phases and requires a specific
correlation between $\varphi_{\mu}$ and  $\varphi_1=\varphi_3$ as shown
in Figure \ref{figsix}. To demonstrate the coincidence of the regions allowed by the 
experimental constraints on the EDM's, we 
choose $m_{3/2}=150\,\rm{GeV}$, $\theta=0.4$, and $\tan\beta=2$, which
leads to a reasonably light superpartner spectrum.
In Fig. \ref{figsixbar}, we plot the
allowed regions for both electron and neutron EDM depending on the
values of $\Theta_1$ and $\Theta_2=\sqrt{1-\Theta_1^2}$ while $\Theta_3$
is set to zero. Frame {\it a)}, where $\Theta_1=0.85$, shows a very
precise  overlap between the electron and neutron EDM allowed regions. 
In frame {\it b)}, 
the magnitudes of all three gaugino
masses are equal but have different
phases due to the different origin of $M_1$ and $M_3$ compared to $M_2$.
Finally, in frame {\it c)} we set  $\Theta_1=0.55$ 
so the magnitude of $M_2$ is significantly larger than 
that of $M_1=M_3$; in this case the alignment between
the EDM allowed regions is spoiled and only small CP- violating phases
are allowed. 

\begin{figure}[h!]
\centering
\epsfxsize=5.25in
\hspace*{0in}
\epsffile{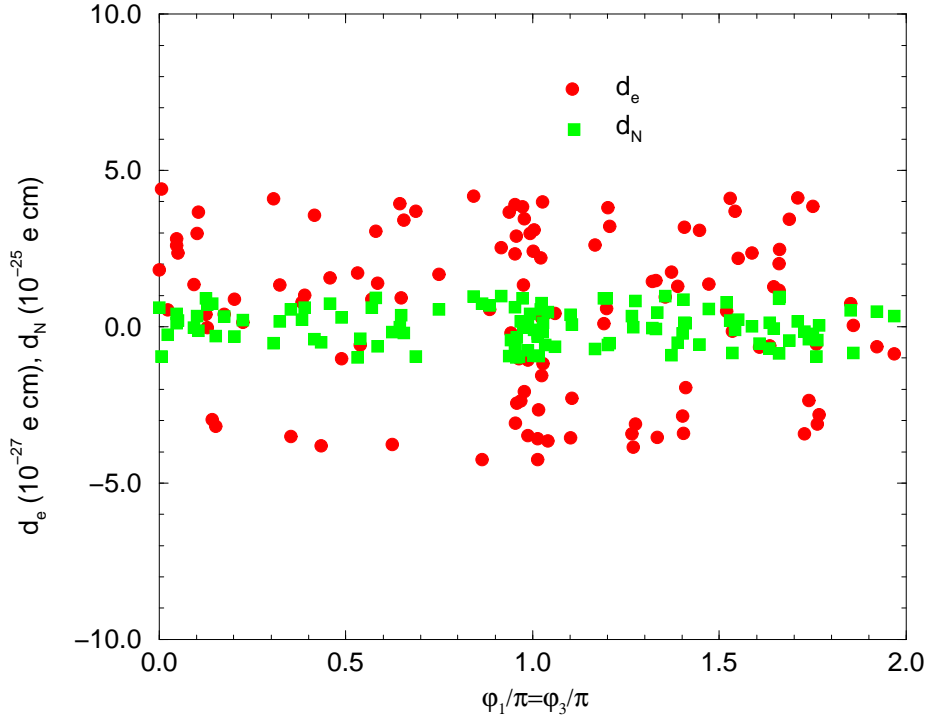}
\bigskip
\caption{Range of the electron and neutron EDM values vs.
$\varphi_1=\varphi_3$ predicted by Eqs. (\ref{model3}) and (\ref{model4})
for the parameters of Figure \ref{figsix}{\it a)}.  All of the points are
allowed by the experimental bounds on the EDM's (note the different scales
for the
eEDM and nEDM).}
\label{figseven}
\end{figure}

It is also interesting to observe that the actual
values of the electron and neutron EDM's for the allowed  points in the
phase parameter space are typically slightly below the experimental limit
and should be within the reach of the next generation of EDM measuring
experiments. In Figure \ref{figseven} we plot the EDM values for the
allowed points in the case of $\Theta_1=0.85$ with all the other
parameters set to
the same values as in previous discussion of Figure \ref{figsix}. This
indicates that if the  CP- violating phases indeed originate from this
type of D- brane configuration, non-zero measured values for both EDM's
much bigger than the SM prediction can be expected.

However, the other orientifold model (in which $SU(2)$ and $U(1)_Y$ arise
from the $5_1$ brane sector) does not allow for large phase solutions.
The reasons for this behavior are similar to that of the
Ho\v{r}ava-Witten scenario: we can use the $U(1)_R$ symmetry of the soft
terms to put $\varphi_2=\varphi_1=0$, which severely limits the
possibility of cancellation between the chargino and neutralino
contributions to the electron EDM. The effect of $\varphi_{A_e}$ alone is
not enough to offset the potentially large chargino contribution and only
a very narrow range of values of $\varphi_{\mu}$ (close to $0$,
$\pi$,$\ldots$) passes the electron EDM constraint. Hence, except at
isolated points in the parameter space of this model, the phases must
be at or below the traditional bound $\simlt {\cal O}(10^{-2})$ to satisfy
the EDM constraints without assuming large sparticle masses.

\section{Conclusions}
\label{VI}
In this paper, we have investigated the possibility that the soft breaking
terms derived in classes of superstring models have large CP- violating
phases that satisfy the phenomenological bounds on the electric dipole
moments of the electron and neutron through cancellations.  The analysis
builds on the work of \cite{nath} and \cite{bgk}, who demonstrated that
this effect can allow for viable points in the MSSM parameter space with
large phases and light superpartner masses, providing for an alternate
resolution to the supersymmetric CP problem.

Sufficient cancellations among the contributions to the EDM's
are difficult to achieve unless there are large relative phases in the
gaugino mass parameters \cite{bgk}. This feature strongly depends on
the string model under consideration; for example, large phases consistent
with the EDM constraints are disfavored in perturbative heterotic string
models and models based on Ho\v{r}ava-Witten theory (in the overall
modulus limit), as the gaugino masses are universal at tree level.
However, our analysis demonstrated that this scenario can be achieved
naturally within Type I string models, where the tree-level gaugino masses
may be nonuniversal depending on the embedding of the SM gauge group into
the D- brane sectors. 

We found that within Type I string models in which $SU(3)$ and $U(1)_Y$
arise from one five-brane sector and $SU(2)$ arises from another set of
five-branes, the cancellations among different contributions to the EDM's
occur over a remarkably wide range of parameter space. In this scenario,
the typical values of the electric dipole moments are not much smaller
than the current experimental limits. Equally remarkably, if we alter the
SM embedding such that $SU(2)$ and $U(1)_Y$ arise from the same set of
branes, the EDM constraints exclude large phase solutions.

The results presented in this paper illustrate that large soft
phases are at least consistent with, and perhaps motivated by, some string
models.  Most importantly, the analysis demonstrates how we may be able to
learn about (even nonperturbative) Planck scale physics using low energy
data.  For example, if the phases of the soft breaking parameters are
determined from collider superpartner data, or measured at B factories,
and found to be large, we have seen that they may provide guidance as to
how the SM is to be obtained from four-dimensional compactifications of
string theory.
  
\acknowledgments
L. E. thanks M. Cveti\v{c} for many informative discussions and
suggestions, and for helpful comments on the manuscript.  This work was
supported in part by the U. S. Department of Energy contract
DE-AC02-76CH03000.    

\def\B#1#2#3{\/ {\bf B#1} (19#2) #3}
\def\NPB#1#2#3{{\it Nucl.\ Phys.}\/ {\bf B#1} (19#2) #3}
\def\PLB#1#2#3{{\it Phys.\ Lett.}\/ {\bf B#1} (19#2) #3}
\def\PRD#1#2#3{{\it Phys.\ Rev.}\/ {\bf D#1} (19#2) #3}
\def\PRL#1#2#3{{\it Phys.\ Rev.\ Lett.}\/ {\bf #1} (19#2) #3}
\def\PRT#1#2#3{{\it Phys.\ Rep.}\/ {\bf#1} (19#2) #3}
\def\MODA#1#2#3{{\it Mod.\ Phys.\ Lett.}\/ {\bf A#1} (19#2) #3}
\def\IJMP#1#2#3{{\it Int.\ J.\ Mod.\ Phys.}\/ {\bf A#1} (19#2) #3}
\def\nuvc#1#2#3{{\it Nuovo Cimento}\/ {\bf #1A} (#2) #3}
\def\RPP#1#2#3{{\it Rept.\ Prog.\ Phys.}\/ {\bf #1} (19#2) #3}
\def\etal{{\it et al\/}}

%
%

\bibliographystyle{prsty}

\begin{thebibliography}{99}
%
\bibitem{dimopoulos}{S. Dimopoulos and D. Sutter, \NPB{452}{95}{496}.}
%
\bibitem{oldedm}{J. Ellis, S. Ferrara, and D.V. Nanopoulos,
\PLB{114}{82}{231}; W. Buchm\"uller and D. Wyler, \PLB{121}{83}{321};
J. Polchinski and M. Wise, \PLB{125}{83}{393}; F. del Aguila, M. Gavela,
J. Grifols, and A. Mendez, \PLB{126}{83}{71}; D.V. Nanopoulos and
M. Srednicki, \PLB{128}{83}{61}.}
%
\bibitem{edm}{M. Dugan, B. Grinstein, and L. Hall, \NPB{255}{85}{413}.}
%
\bibitem{gar}{R. Garisto, \NPB{419}{94}{279}}
%
\bibitem{nath}{T. Ibrahim and P. Nath, \PRD{58}{98}{111301};
\PRD{57}{98}{478}, Erratum-ibid \PRD{58}{98}{019901}; \PLB{418}{98}{98}.}
%
\bibitem{bgk}{M. Brhlik, G. Good, and G. L. Kane, hep-ph/9810457.}
%
%
\bibitem{pokorski}{S. Pokorski, J. Rosiek, and C. Savoy, hep-ph/9906206.}
%
\bibitem{bk}{M. Brhlik and G. L. Kane, \PLB{437}{98}{331}.}
%
\bibitem{dine}{M. Dine, R. Leigh, and D. MacIntire,
\PRL{69}{92}{2030};
K. Choi, D. Kaplan, and A. Nelson, \NPB{391}{93}{515}.}
%
\bibitem{bim}{A. Brignole, L. Ib\'a\~nez, and C. Mu\~noz,
\NPB{422}{94}{125}, Erratum-ibid \NPB{436}{95}{747}.}
%
\bibitem{bims}{ A. Brignole, L.
Ib\'a\~nez, C. Mu\~noz, and C. Scheich, Z. Phys. {\bf C74},
157 (1997), hep-ph/9508258.} 
%
\bibitem{bimrev}{A. Brignole, L. Ib\'a\~nez, and C. Mu\~noz,
hep-ph/9707209, in {\it Perspectives on Supersymmetry}, ed. G. L. Kane,
World Scientific, 1998; C. Mu\~noz, hep-th/9507108.}
%
\bibitem{orbifolds}{L. Dixon, J.A. Harvey, C. Vafa and E. Witten,
\NPB{261}{85}{678} and \NPB{274}{86}{285}; L. Dixon, D. Friedan, E.
Martinec and S. Shenker, \NPB{282}{87}{13};
L. Ib\'a\~nez and D. L\"{u}st,
\NPB{382}{92}{305};
L.~Ib\'a\~nez, J.E.~Kim, H.P.~Nilles and F.~Quevedo,
\PLB{191}{87}{282};
J.A.~Casas and C.~Mu\~noz, \PLB{209}{88}{214} and \B{214}{88}{157};
J.A.~Casas, E.~Katehou and C.~Mu\~noz, \NPB{317}{89}{171};
A.~Font, L.~Ib\'a\~nez, H.P.~Nilles and F.~Quevedo,
\PLB{210}{88}{101};
A.~Chamseddine and M.~Quir\'os, \PLB{212}{88}{343},
\NPB{316}{89}{101};
A.~Font, L.~Ib\'a\~nez, F.~Quevedo and A.~Sierra,
\NPB{331}{90}{421}; M. Cveti\v{c}, \PRL{59}{87}{1795},
\PRL{59}{87}{2829}.}
%


\bibitem{horavawitten}{P. Ho\v{r}ava and E. Witten, \NPB{460}{96}{506},
\NPB{475}{475}{96}{94}; E. Witten, \NPB{471}{96}{135}, P. Ho\v{r}ava,
\PRD{54}{96}{7561}.}
%
\bibitem{ovrut}{A. Lukas, B. A. Ovrut, and D. Waldram, JHEP 9904:009
(1999),  \PRD{59}{99}{106005}; R. Donagi, A. Lukas, B. A. Ovrut, and D.
Waldram, hep-th/9811168.}
%
\bibitem{nilles}{H. P. Nilles, M. Olechowski, and M. Yamaguchi,
\PLB{415}{97}{24}, \NPB{530}{98}{43}; K. Meissner, H. P. Nilles, and M.
Olechowski, hep-th/9905139; D. Bailin, G. V. Kraniotis, and A.
Love, hep-ph/9906507, \PLB{432}{98}{90}; Z. Lalak and S. Thomas,
\NPB{515}{98}{55}.}
%
\bibitem{munoz}{ K. Choi, H. B. Kim, and C. Mu\~noz,
\PRD{57}{98}{7521}; D. G. Cerde\~no and C. Mu\~noz, hep-ph/9904444; T. Li,
hep-th/9908013, \PRD{59}{99}{107902}, \PRD{57}{98}{7539}; T. Kobayashi,
J. Kubo, and H. Shimabukuro, hep-ph/9904201.}
%
\bibitem{barrkhalil}{S. Barr and S. Khalil, hep-ph/9903425.}
%

\bibitem{sagnotti}{G. Pradisi and A. Sagnotti, \PLB{216}{89}{59}; M.
Bianchi and A. Sagnotti, \PLB{247}{90}{517}, \NPB{361}{91}{519}; E. Gimon
and J. Polchinski, \NPB{477}{96}{715},
hep-th/9601038; A. Dabholkar and J. Park, \NPB{477}{96}{207},
hep-th/9602030;
\NPB{477}{96}{701}, hep-th/9604178; \PLB{394}{97}{302}, hep-th/9607041;
E. Gimon and C. Johnson, \NPB{477}{96}{715}, hep-th/9604129;
J. Blum and A. Zaffaroni, \PLB{387}{96}{71}, hep-th/9607019;
J. Blum, \NPB{486}{97}{34}, hep-th/9608053;  
C. Angelantonj, M. Bianchi, G. Pradisi, A. Sagnotti and Ya.S. Stanev,
\PLB{385}{96}{96}, hep-th/9606169; C. Angelantonj, M. Bianchi,
G. Pradisi, A. Sagnotti, and Ya.S. Stanev, \PLB{387}{96}{743}. }
%
\bibitem{kakushadze}{Z. Kakushadze, G. Shiu and S.-H. Tye,
\PRD{58}{98}{086001}, hep-th/9803141; M. Berkooz and R.G. Leigh,
\NPB{483}{97}{187},
hep-th/9605049; G. Zwart, \NPB{526}{98}{378}, hep-th/9708040; Z.
Kakushadze,
\NPB{512}{98}{221}, hep-th/9704059;
Z. Kakushadze and G. Shiu, \PRD{56}{97}{3686}, hep-th/9705163;
\NPB{520}{98}{75}, hep-th/9706051;
L.E. Ibanez, JHEP 9807:002 (1998); D. O'Driscoll, hep-th/9801114;
J. Lykken, E. Poppitz, and S. Trivedi, \NPB{543}{99}{105},
hep-th/9806080; M. Cveti\v{c}, L. Everett, P. Langacker, and J. Wang,
JHEP 9904:020 (1999), hep-th/9903051.}
%
\bibitem{ibanez}{L. Ib\'a\~nez, C. Mu\~noz, and S. Rigolin,
hep-ph/9812397.}
%
\bibitem{shiu}{G. Shiu and S.-H. Tye, \PRD{58}{98}{106007},
hep-th/9805157.}
%
\bibitem{nexpnew}{P. G. Harris, et al, \PRL{82}{99}{904}.} 

\bibitem{eexp} {E.D. Commins, S.B. Ross, D. DeMille, and B.S. Regan, 
{\it Phys. Rev.} {\bf A50}, 2960 (1994);
K. Abdullah et al., \PRL{65}{90}{2347}.}
%
\bibitem{thomas}{S. Dimopoulos and S. Thomas, \NPB{465}{96}{23}.}
%
\bibitem{wagner}{A. Pilaftsis and C.E.M. Wagner, hep-ph/9902371; D. A.
Demir, hep-ph/9901389.}
%
\bibitem{reparameterization}{R. Akhoury, M. Brhlik, G. L. Kane, and
A. Sinkovics, in preparation.}
%
\bibitem{muproblem}{J. Kim and H. P. Nilles, \PLB{138}{84}{150}; G.
Giudice and A. Masiero, \PLB{206}{88}{480}.}
%
\bibitem{orbimu}{A. Brignole, L. Ib\'a\~nez, and C. Mu\~noz,
\PLB{387}{96}{769}; Y. Kawamura, T. Kobayashi, and M. Watanabe,
\PLB{419}{98}{157}; C. Kokorelis, hep-th/9810187.}


\bibitem{quevedo}{See e.g. J. Lykken, hep-ph/9511456; F. Quevedo,
hep-th/9603074.}
%
%
\bibitem{choi}{K. Choi, \PRL{72}{94}{1592}.}
%
%
\bibitem{decarlos}{B. de Carlos, A. Casas, and C. Mu\~noz,
\PLB{299}{93}{234}.}
%
\bibitem{tduality}{A. Font, L. Ib\'a\~nez, D. L\"{u}st, and F. Qeuevedo,
\PLB{245}{90}{401}.}
%
\bibitem{cvetic}{M. Cveti\v{c}, A. Font, L. Ib\'a\~nez, D. L\"{u}st, and
F. Quevedo, \NPB{361}{91}{232}.}
%
%
\bibitem{bailin1}{B. Acharya, D. Bailin, A. Love, W. A. Sabra, and S.
Thomas, \PLB{357}{95}{387}.}
%
\bibitem{bailin2}{D. Bailin, G. V. Kraniotis, and A. Love,
\PLB{414}{97}{269}; \NPB{518}{98}{92}.}
%
\bibitem{kaplunovsky}{V. Kaplunovsky, \NPB{307}{88}{145}, Erratum-ibid
\NPB{382}{92}{436}, hep-th/9205070.}
%
\bibitem{dienes}{K. Dienes and A. Faraggi, \NPB{457}{95}{409};
        K. Dienes, A. Faraggi and J. March-Russell,
\NPB{467}{96}{44}.}
%
\bibitem{gunion}{C. Chen, M. Drees, and J. Gunion, \PRD{55}{97}{330}.}
%
\bibitem{casas}{A. Casas, A. Ibarra, and C. Mu\~noz, hep-ph/9810266.}
%
\bibitem{graham} {G.D. Kribs \NPB{535}{98}{41}.}
%
\bibitem{sugracp}{R. Garisto and J. D. Wells, \PRD{55}{97}{1611}}.
%
\bibitem{conif}{A. Uranga,  JHEP 9901:022 (1999).}
%
\bibitem{typeIanom}{M. Berkooz, R. G. Leigh, J. Polchinski, J.H. Schwarz,
N. Seiberg and E. Witten, \NPB{475}{96}{115}; L.E. Ibanez, R. Rabadan and
A.M. Uranga, \NPB{542}{99}{112}; E. Poppitz, \NPB{542}{99}{31}.}
%
\bibitem{rad}{ L. Ib\'a\~nez and G. Ross, \PLB{110}{82}{215}.}
%
%
%
\bibitem{pilaftsis}{D. Chang, W. Keung and A. Pilaftsis,
\PRL{82}{99}{900}. }

\end{thebibliography}

\end{document}